\documentstyle[epsf]{elsart}

\begin{document}
\begin{frontmatter}
\title{Pions in isospin asymmetric matter and nuclear Drell-Yan
scattering}
\author{C.~L.~Korpa}
\address{Department of Theoretical Physics, Janus Pannonius
University, Ifjusag u.\ 6, 7624 Pecs, Hungary}
\author{A.~E.~L.~Dieperink}
\address{Kernfysisch Versneller Instituut, Zernikelaan 25, NL-9747AA
Groningen, The Netherlands}

\begin{abstract}
Using a self-consistent delta-hole model the pion propagation
in isospin asymmetric nuclear matter is studied. In neutron-rich 
matter, corresponding to heavy nuclei, a significant difference in
positive and negative pion light-cone distributions is obtained 
leading to a
nuclear enhancement of up antiquark distribution compared
to the down antiquark one. This means that the sea-quark
asymmetry in the free nucleon cannot be extracted model
independently from an experiment on a neutron-rich
nucleus.
\end{abstract}

{\it PACS:} 21.65.+f; 24.85.+p; 13.40.-f

\begin{keyword}
Isospin-asymmetric medium; Drell-Yan scattering; 
Antiquark flavour asymmetry.
\end{keyword}

\end{frontmatter}

\section{Introduction}
The meson-cloud model plays an important role in dealing with
non-perturbative Quantum Chromodynamics effects in the nucleon.
It has been used by several groups to interpret momentum
distributions
of sea quarks in the nucleon
\cite{Thomas83,Hwang91,Kumano91,Szczurek96,Holtmann96},
measured in deep inelastic scattering.
The standard approach is based upon the Sullivan process
\cite{Sullivan72}, in which
the only essential parameter is the cut-off in the pion-nucleon-nucleon
vertex.

Originally the emphasis was mainly on the description of the isoscalar
$\bar{u}+
\bar{d}$ distributions.
More recently, since the observed violation of the Gottfried sum rule,
showing an excess of $\bar{d}$ over $\bar{u}$
in proton, the interest focussed on the
properties of $\bar{u}(x)-\bar{d}(x)$ in the
nucleon,
whose $x$ dependence was measured recently \cite{Hawker98}. 
The asymmetry in antiquark distributions has been
interpreted mostly in terms of the pion-cloud model
(a review is given in ref.\ \cite{Kumano98}, see also  
\cite{Melnitchouk98}), and also in a soliton model in 
the large-$N_c$ limit \cite{Dressler98}.

Since pion properties are strongly affected by the nuclear
medium,
the pion cloud plays also an important role in modelling
nuclear effects on deep inelastic processes. It was used
in the past in connection with the EMC effect 
\cite{Ericson83,Jung88,Marco96},
leading to a few percent enhancement of the structure-functions ratio
around $x=0.2$.
Similarly, most approaches predicted a nuclear enhancement of
the pion field and hence 
the $\bar{u}(x)$ and $\bar{d}(x)$ distributions, leading
to a noticeable increase in the predicted Drell-Yan (DY) 
cross-section ratio.
On the other hand experimentally the  DY scattering
on  nuclear targets \cite{Alde90}
did not show much evidence for enhancement over nucleon targets.
Several papers have dealt with this discrepancy, pointing out
different mechanisms leading to reconciliation with the
measurements \cite{Szczurek96,Holtmann96,Brown95,Dieperink97}.

For the purpose of presenting our results we find it convenient
to separate them into two aspects. The calculation of experimentally
measured proton-nucleus to proton-deuteron DY cross-section ratio
is our first aim.
The isospin
asymmetric medium affects the various isospin states of the
nucleon's pion cloud differently, leading to an excess of
up antiquarks over down ones (even if the distributions in
free proton are identical) for nuclei with more neutrons
than protons. Since the charges of up and down quarks are
different, the effect also shows up in the nuclear DY
scattering cross section.
In addition to presenting the DY ratio, one can also
investigate possible nuclear effects on the difference
$\bar{u}(x)-\bar{d}(x)$ for a nucleon in the medium.
It has been pointed out by Kumano that in a
neutron-excess nucleus there could be medium effects
contributing to this
difference, coming from flavor-asymmetric parton recombination in the 
small
Bjorken-$x$ region \cite{Kumano95}. The mechanism he considered
contributes only at relatively small $x$ values ($x<0.1$) and the
effect is also rather small (2--10\% of the observed asymmetric
sea in the nucleon).

In the present calculation we extend the effective field theory approach
to compute pion properties in isospin-symmetric nuclear medium
\cite{Korpa95} to the case of arbitrary proton and neutron
densities. Then we use the pion-cloud model \cite{Dieperink97} to
compute the medium-modified quark and antiquark distributions
and the proton-nucleus DY cross section.
We assume that the medium effects solely originate from a 
modification
of the pion cloud. 
\footnote{It is known 
\protect{\cite{Koepf96}} that the pion and
other mesons 
cloud cannot account completely for the antiquark distribution
of the nucleon. For example, gluon splitting gives
a sizeable contribution at small $x$ and large $Q^2$, but
this contribution is approximately flavor symmetric 
\protect{\cite{Koepf96}} and thus should not modify our results 
significantly.
} 
The nucleons are treated in mean field approximation, while the
delta-isobar and the pion are dressed self-consistently. The
effects of short-range baryon repulsion are included through
Migdal's $g'$ parameters ($g'_{NN},g'_{N\Delta},
g'_{\Delta\Delta}$). Sensitivity of the pion distribution to
the delta-hole self-energy was
already recognized in ref.\ \cite{Ericson83}, thus we carried out
the computation with the self-consistently
determined in-medium delta-isobar spectral function \cite{Korpa95}.

We compute the pion light-cone momentum distributions
separately for the three charge states as a function of the asymmetry
parameter $\beta\equiv (N-Z)/A$. For $\beta>0$ the $\pi^-$ distribution
exceeds that of $\pi^0$, which in turn is larger than the $\pi^+$
distribution, as expected on the basis of particle-hole self-energy
relationships.
\section{Drell-Yan cross section}
The Drell-Yan cross section for the process $p+A \rightarrow 
\mu^+\mu^-X$
is given by (suppressing the $Q^2$ dependence)
\begin{equation}
d^2\sigma= \frac{4\pi \alpha K(x_1,x_2)}{ 9s x_1x_2}
\sum_f e^2_f[ q_f(x_1) \bar{q}_f(x_2)
   +\bar{q}_f(x_1)q_f(x_2)] dx_1dx_2 ,
\end{equation}
where the sum is over all flavors, and $x_1, \ x_2$ are the 
longitudinal
momentum fractions carried by quark of the beam and target nucleons, 
respectively.
By a suitable selection of kinematics
the values of $x_1, x_2$ can be deduced from experiment \cite{Alde90}.

If we consider the region $x_1>0.3$ when the antiquarks in the
projectile play
a negligible role, the ratio of the proton-nucleus to proton-deuteron
DY cross-sections takes on the form
\begin{eqnarray}
R_{Ad} &\equiv &
\frac{2}{A} \frac{d\sigma(pA)}{d\sigma(pd)}
=\frac{\bar{u}_{N/A}(x_2)+\bar{d}_{N/A}(x_2)}
{\bar{u}_p(x_2)+\bar{d}_p(x_2)} \nonumber \\
&+&f(x_1)
\frac{\bar{u}_{N/A}(x_2)-\bar{d}_{N/A}(x_2)}
{\bar{u}_p(x_2)+\bar{d}_p(x_2)}, \label{dyratio1}
\end{eqnarray}
where $f(x_1)\equiv (4u_p(x_1)-d_p(x_1))/(4u_p(x_1)+d_p(x_1))$
is close to unity. Here $\bar u_p$ and $\bar d_p$ are antiquark
distributions in the free proton, while 
$\bar{u}_{N/A}$ and $\bar{d}_{N/A}$ are the antiquark 
distributions per nucleon in the nucleus, differing from the
free-nucleon distributions by
the medium modified pion-cloud contribution. Denoting the latter
by $\delta \bar{u}_{\pi/A}$ and $\delta \bar{d}_{\pi/A}$ the 
DY ratio becomes
\begin{eqnarray}
R_{Ad}&=& 1 +
\frac{1}{\bar u _p (x_2)+\bar d _p(x_2)} \left\{
\delta \bar{u}_{\pi/A}(x_2)+\delta \bar{d}_{\pi/A}(x_2)
\right. \nonumber \\
&+&
\left.
\beta f(x_1)
\left[
\bar d _p(x_2)-\bar u _p(x_2)
+
\left( \delta \bar{u}_{\pi/A}(x_2)-\delta \bar{d}_{\pi/A}(x_2)
\right)/\beta 
\right]  \right\}.  \label{dyratio2}
\end{eqnarray}
We see from the above expression that apart from nuclear effects,
leading to nonzero $\delta \bar u_{\pi/A}$ and
$\delta \bar d_{\pi/A}$, for $\beta\neq 0$ there is a nucleonic
one \cite{Szczurek96}, stemming from the nonzero value of the 
antiquark-distribution difference $\bar d_p-\bar u_p$ in the 
free proton.
This underlines the necessity to use parton
distributions in accordance with latest $\bar d _p - \bar u_p$
observations (for discussion on this point see section IV).
For the relatively small asymmetries of interest for stable nuclei
the ratio $(\delta \bar{u}_{\pi/A}-\delta \bar{d}_{\pi/A})/\beta$
appearing in the above expression is practically independent of 
$\beta$, leading to a linear dependence of $R_{Ad}$ on it.


For the change of antiquark distributions due to pion-cloud 
modification we use the convolution formula \cite{Dieperink97}
$$ \delta \bar{q}_{f,\pi/A}(x)=
\sum_a \int_x^A \frac{dy}{y} \delta f_\beta^{\pi^a/A} (y) \bar{q}
_f^{\pi^a}(x/y), $$
where $\delta f_\beta^{\pi^a}(y)$, given by
\begin{equation}
\delta f^{\pi^a}_\beta (y)
=\frac{1+\beta}{2} \left( f^{\pi^a/n/A}(y)-f^{\pi^a/n}(y)\right)
+\frac{1-\beta}{2} \left( f^{\pi^a/p/A}(y)-f^{\pi^a/p}(y)\right),
\end{equation}
represents the change of the pion light-cone-momentum
distribution per nucleon in the medium.
$f^{\pi^a/n/A}$ ($f^{\pi^a/p/A}$) and
$f^{\pi^a/n}$ ($f^{\pi^a/p}$) denote the distribution of $\pi^a$ per
neutron (proton) in medium and in free space, respectively.
They are discussed in the next section.
\section{Pions in isospin asymmetric nuclear medium}
We consider a model consisting of pions, nucleons and 
delta-isobars in an
infinite, spatially uniform system at zero temperature. The
proton and neutron densities are given through their chemical
potentials $\mu_p$ and $\mu_n$. The equilibrium conditions 
for nucleons, delta-isobars and pions 
imply
that the chemical potential for the neutral pion is zero, while
those of charged pions are: $\mu_{\pi^-}=-\mu_{\pi^+}=
\mu_n-\mu_p$. The antiparticles of nucleons and isobars are neglected,
but a relativistic kinematics is used. The nucleons are further
treated in the mean-field approximation, with 
momentum-independent mass ($M_{*p}-M_p,M_{*n}-M_n$) and energy shifts
($c_p,c_n$) 
modelling their binding, i.e.\ $E_n(p)=\sqrt{M_{*n}^2+p^2}+c_n$ for
neutrons and
$E_p(p)=\sqrt{M_{*p}^2+p^2}+c_p$ for protons.
The Schwinger-Dyson equations without
vertex corrections are then solved self-consistently for the 
delta-isobar
and pion \cite{Korpa95}. The pion self-energy consists of the
particle-hole and delta-hole contributions, with both imaginary
and real parts taken into account, assuring correct analytical
properties.
Short-range baryon repulsion is taken into account through
Migdal's $g'_{NN},g'_{N\Delta},g'_{\Delta\Delta}$ parameters.
The sum-rules for spectral functions of pions and deltas
are checked and found to be satisfied to 1--2\%.

The pion light-cone momentum distributions are then calculated, using
the in-medium pion self-energy.
Direct calculation of the diagrams corresponding to the pion
emitted by the in-medium nucleon proceeds analogously to
ref.\ \cite{Dieperink97}, but separately for the three charge states
of the pion and taking into account different
neutron and proton densities (and thus different $M_{*p},M_{*n}$
effective masses and $c_p,c_n$ energy shifts).

It is well known that $f(y)$ can be expressed compactly in terms
of an integral over the spin-isospin response function or
imaginary part of the pion propagator. It is also possible to compute it
directly from summing contributing diagrams \cite{Dieperink97},
which represent emission of a dressed pion by a nucleon. 
The latter procedure we prefer for
numerical reasons, since it needs computation of the pion self-energy
in smaller energy-momentum region. The 
schematic expression for the pion light-cone distribution per nucleon 
has the form
    $$f_\beta^{\pi/p/A}(y)= \frac{2}{\rho}
\int \frac{d^3p}{(2\pi)^3} 
\int \frac{d^3 p'}{(2\pi)^3 2E(p')}
  \delta ( y- \frac{k^0+k^3}{M}) |X^{\pi}_\beta(k)|^2, $$
where $X^\pi_\beta$ represents the sum of relevant diagrams,
$\rho$ the nucleon density, 
$k_0=E(p)-E(p'), \vec k=\vec p -\vec{p}\,'$ and $\vec p$ 
($\vec{p}\,'$) is 
the momentum of
incoming (outgoing) nucleon.

The full expression generalizing the case for isospin-symmetric medium from
\cite{Dieperink97} reads (for the special case of $\pi^+$ on proton):
\begin{eqnarray}
f^{\pi^+/p/A}_\beta(y)&=&\frac{3y
g^2_{\pi NN} M_*} {2(2\pi
p_{Fp})^3M}
\int_{-p_{Fp}}^{p_{Fp}} dp_3 \int_0^{\sqrt{p^2_{Fp}-p_3^2}}
p_\perp dp_\perp
\int_{p^{\prime}_{\perp \mbox{\scriptsize{min}}} }^\infty
p'_\perp dp'_\perp \nonumber \\
 && \times\int_0^{2 \pi} d\theta k^2 F_{\pi NN}^2(k) {1\over z'}
\,|\tilde{D}_{\pi^+}(k_0,k)|^2,
\label{pidis}
\end{eqnarray}
where $p_{Fp}$ is the Fermi momentum of the proton,
$M$ the nucleon's mass, 
$M_*\equiv (M_{*p}+M_{*n})/2$,
$g_{\pi NN}$ the $\pi^0 NN$
coupling
and $F_{\pi NN}(k)$ the form factor,
while $\theta$ is the angle between $\vec{p}_\perp$ and
$\vec{p}\,'_\perp$. 
$\tilde{D}_{\pi^+}$ is
the full pion propagator, corrected for the presence of four-fermion
couplings through Migdal's $g'$ parameters, given as $\tilde{D}
_{\pi N}$ in ref.\ \cite{Dieperink97},
$p^\prime_{\perp \mbox{\scriptsize{min}}}=(2z'M_*
(M_{*n}^2+p_{Fn}^2)^{1/2}-
(z'^2+1)M^2_{*p})^{1/2}$ if the argument of the square root is positive,
otherwise is zero, and
$
z' \equiv (-M y+p_3+E(p)-c_n)/M_*. 
$
The expression for $\pi^-$ is obtained by swapping the indices 
$n$ and $p$ and
for neutral pions the total distribution is a sum of
two terms corresponding to emission by proton or neutron.

The difference in distributions for the various charge states of
pion basically comes from two factors. One is the Pauli blocking
of the outgoing nucleon, which in neutron-rich medium 
restricts emission of $\pi^+$ (from
a proton, creating a neutron in the final state) more than the
emission of $\pi^-$ (since a proton appears in this case in the 
final state). The other effect is the dressing of the pion 
propagator, in which the particle-hole and delta-hole self-energies
enter. Since the dominant contribution comes from
the particle-hole contribution for $N>Z$ 
(neutron density larger than the proton one) 
the $\pi^-$ propagator is affected
more than the one of the $\pi^+$ 
(more details are given in the next section).

While the delta has been shown to play an important role in the
asymmetry for free nucleons, the medium effects are negligible 
\cite{Dieperink97} and not included here, which is partially a 
consequence of the use of a soft pion-nucleon-delta form-factor,
as obtained from a fit 
to pion-nucleon scattering \cite{Korpa95}. 
Since the isobar's contribution for $\beta\neq 0$ to
$\delta \bar{u}_\pi-\delta \bar{d}_\pi$ is of the opposite sign
compared to the contribution of Eq.\ (\ref{pidis}), as discussed 
in the next section, at small $x$ 
inclusion of this term might result in a small decrease of the
calculated isospin-asymmetry effect.

\section{Results and discussion}
For numerical calculation we used the proton to neutron ratio of
tungsten, for which there are measurements of the DY cross
section \cite{Alde90}. The asymmetry parameter in this case 
is $\beta=0.196$ and
for the Fermi momenta of protons and neutrons we chose
$p_{Fp}=238\;$MeV and $p_{Fn}=272\;$MeV, corresponding to total
nucleon density slightly below the saturation density. To take into
account the different binding of protons and neutrons for the energy
shifts we take $c_p=40\;$MeV, $c_n=42\;$MeV, with the common
effective mass $M_{*n}=M_{*p}=0.85\;$GeV, thus assuring the correct
asymmetry energy of 28 MeV.
For the pion-nucleon-nucleon vertex we use a dipole form-factor with
cut-off $\Lambda=1\;$GeV. We checked that varying the cut-off in the
range $0.9 - 1.1$ GeV does not change appreciably the medium effect
on the pion.

In Fig.\ 1 we present the pion light-cone-momentum
distributions for nucleons in medium and in free space (upper
four curves), as well as the excess pion distributions (lower
three curves). The $\pi^+$
($\pi^-$) distributions are per proton (neutron), while that of
$\pi^0$ is per average nucleon in the medium. 
The neutral pions see little difference between an isospin-symmetric
and an isospin-asymmetric nuclear medium, as long as the total
nucleon density is the same (actually, the different mean-field
shifts for proton and neutron may lead to a very small effect).
However, for the case of neutron excess, the $\pi^-$ distribution
per neutron is larger than the $\pi^+$ distribution per proton.
This comes from a difference in particle-hole self energies, which
is mainly responsible for the light-cone-momentum distributions.
It is easy to understand the difference if we look at the imaginary
part of the self energy as a function of pion's energy. If the
energy is positive, the pion can excite a neutron from the
Fermi sea to become
a proton (above the proton Fermi sea) if its charge is positive. 
A negative pion
can make from a proton in the Fermi sea a neutron above the
neutron Fermi sea.
Since there
are more neutrons than protons in the Fermi sea, 
the absolute value of the
imaginary part of the $\pi^+$ self energy will be larger than
that of $\pi^-$. Bearing in mind that in expression
(\ref{pidis}) the pion energy is negative (for the dominating
part of the integraton region) and using the relation that
$
\mbox{Im}\, \Pi_{\pi^+}(\omega,k)=
\mbox{Im}\, \Pi_{\pi^-}(-\omega,k)
$
we arrive at the inequality $f^{\pi^-/n/A}(y)>f^{\pi^+/p/A}(y)$, 
in accordance with the numerical computation.
We mention that for the delta-hole self-energies the relationship
between the $\pi^-$ and $\pi^+$ is
the opposite to that of particle-hole one,
i.e.\ for positive energy that of $\pi^-$ is larger
(in absolute value) than the self-energy of $\pi^+$; a consequence of
different isospin factors in the pion-nucleon-delta vertex.
We see that for the
studied asymmetry there is a significant difference in the
distributions of three charge states. The distribution in isospin
symmetric medium is very close to the $\pi^0$ distribution.

\begin{figure}
\epsfysize=8cm
\centerline{\epsffile{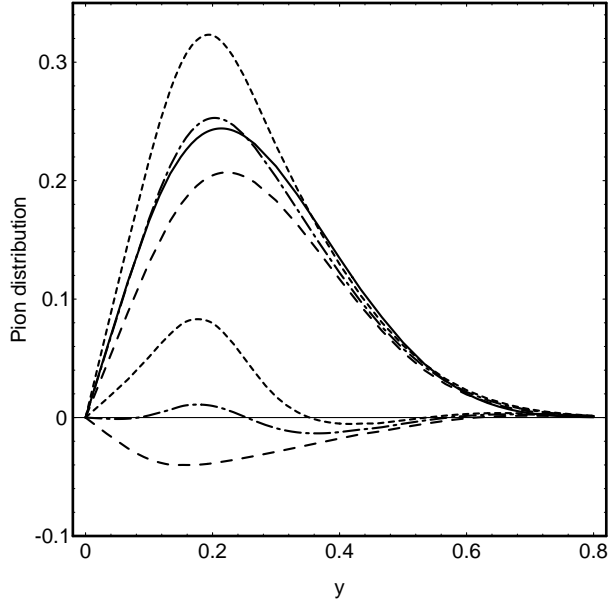}}
\vspace{5mm}
\caption{Pion light-cone-momentum distributions. Full line is for
free nucleon, long dashed for $\pi^+$, short-dashed for $\pi^-$,
dot-dashed for $\pi^0$, with
$g'_{NN}=0.6, g'_{N\Delta}=g'_{\Delta\Delta}=0.3$. The three 
lower curves show the excess distributions with respect to the
free nucleon.
}
\label{fig1}
\end{figure}

The valence distribution of negative pions contains up antiquarks and
thus in
neutron-rich matter they outnumber the down antiquarks present in
the positive pions, due to nuclear effects on the pion clouds.
A comparison of this effect to the recently measured \cite{Hawker98}
$\bar{d}_p(x)-\bar{u}_p(x)$ difference for proton is
presented in Fig.\ 2. 
It shows the quantity 
$(\delta \bar{u}_{\pi/A}-\delta \bar{d}_{\pi/A})/\beta$
compared to the mentioned $\bar d_p(x)-\bar u_p(x)$, to which
it is added in the expression (\ref{dyratio2}) for the DY ratio
$R_{Ad}$. 
The quantity $(\delta \bar{u}_{\pi/A}-\delta \bar{d}_{\pi/A})/\beta$
does not change appreciably with $\beta$ up to its value of 0.2, but
it is sensitive to the $g'$ parameters.
Existing experimental and theoretical information \cite{Brown95}
suggests values of 0.55 -- 0.7 for $g'_{NN}$ and 0.3 -- 0.4 for
$g'_{N\Delta}$ and $g'_{\Delta\Delta}$. These values give results
consistent with observed DY scattering for the isospin-symmetric
calculation \cite{Dieperink97}, and we use these values also in the 
present case. Taking momentum dependent $g'$ parameters (as 
suggested in ref.\ \cite{Brown95}) could change the details of
our results. Exploration of this and other effects 
we leave for a future publication.

\begin{figure}
\epsfysize=8cm
\centerline{\epsffile{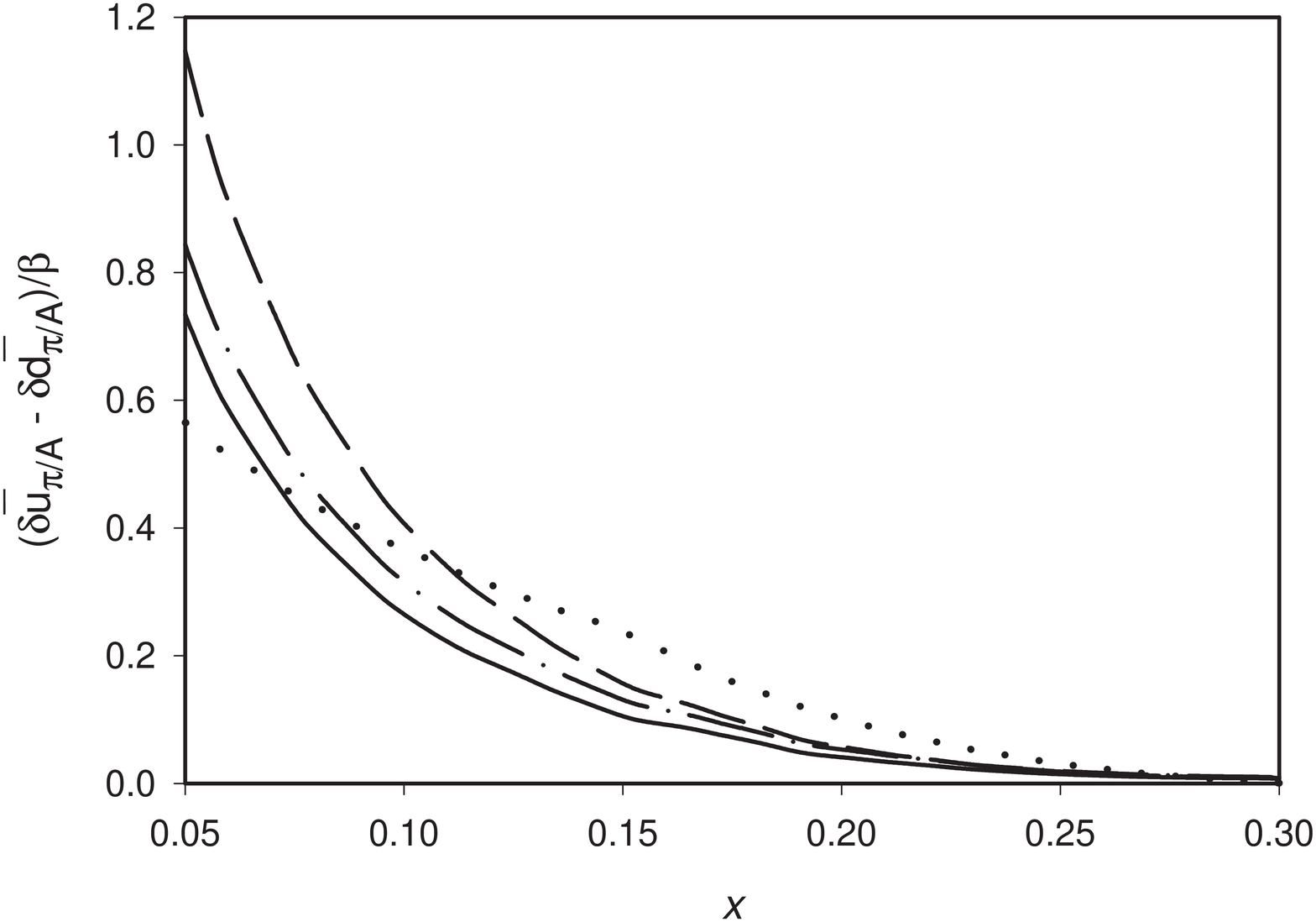}}
\vspace{-17mm}
\caption{The antiquark-distribution difference,
$(\delta \bar{u}_{\pi/A}-\delta \bar{d}_{\pi/A})/\beta$
per nucleon in the medium.
Full line is for
$g'_{NN}=0.6, g'_{N\Delta}=g'_{\Delta\Delta}=0.3$, 
dashed line for
$g'_{NN}=0.5, g'_{N\Delta}=g'_{\Delta\Delta}=0.3$ and
dash-dot line for
$g'_{NN}=0.5, g'_{N\Delta}=g'_{\Delta\Delta}=0.4$.
The dotted line shows $\bar d_p -\bar u_p$ for
comparison.
}
\label{fig2}
\end{figure}

The
free nucleon parton distributions are taken from ref.\ \cite{Lai97}.
They fit the $\bar d_p-\bar u_p$ and $\bar d_p/\bar u_p$ of
ref.\ \cite{Hawker98} quite well up to $x=0.15$. However, at larger
$x$ the measured
difference $\bar d_p - \bar u_p$ and especially the ratio
$\bar d_p/\bar u_p$ are poorly fitted \cite{Hawker98}.
To correct for
this discrepancy, which would cause a significant increase in
calculated $R_{Ad}$,  
we impose a constraint $\bar d_p(x)=\bar u_p(x)$
for $x\geq 0.3$ (by using the arithmetic mean value) 
and interpolate linearly between $x=0.15$, when the
unmodified distributions are used, and the point $x=0.3$.
In this way, the employed distributions fit the measurements 
of ref.\ \cite{Hawker98} very well.
From Fig.\ 2 we see that the two terms, whose sum appears in the
square bracket of Eq.\ (\ref{dyratio2}), are of comparable 
magnitude, i.e.\ the nuclear effect of the isovector part of the
DY ratio plays as important role as the nucleon antiquark asymmetry.

Since the squared charge of up antiquarks is four times that of down
antiquarks, the former enter in the DY cross section with
correspondingly larger weight. This implies an enhancement of the
DY cross-section ratio for neutron-rich medium compared to
the isospin symmetric case.
In Fig.\ 3 the ratio of cross-sections (per nucleon) is shown as
a function of $x_2$, where
both in the numerator and denominator integration over $x_1$ is
performed for $x_1>x_2+0.2$, corresponding to the experimental
situation of ref.\ \cite{Alde90}. The full line (dash-dot line)
corresponds to
$\beta=0.196$ and $g'_{NN}=0.6$ ($g'_{NN}=0.5$), 
dotted line is for $\beta=0$, while the
dashed line is for asymmetric medium ($\beta=0.196$), but
without medium effect on the pion cloud. Measurements from
ref.\ \cite{Alde90} for W are shown as points with error bars.
The errors are too large for any definite
statement to be made on the isospin-asymmetry effect.

\begin{figure}
\epsfysize=8cm
\centerline{\epsffile{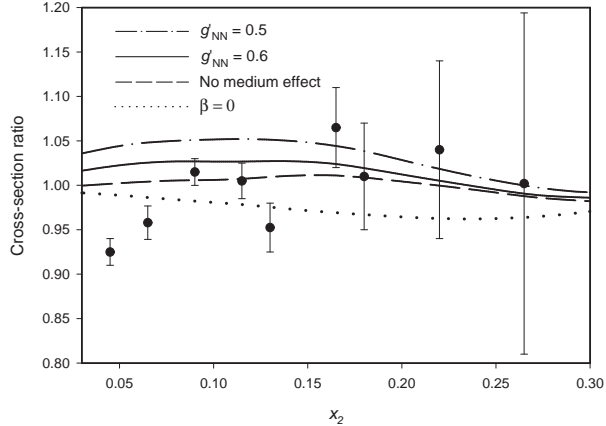}}
\vspace{-16mm}
\caption{DY cross-section ratio. Full (dash-dot) line is for
asymmetric nuclear medium with $\beta=0.196$ and
$g'_{NN}=0.6$
($g'_{NN}=0.5$), 
dotted line
for symmetric matter with $g'_{NN}=0.6$; in both cases
$g'_{N\Delta}=g'_{\Delta\Delta}=0.3$.
The dashed line is for $\beta=0.196$, but without nuclear
pion effects.
Experimental results from ref.\ \protect{\cite{Alde90}} 
for W are shown as points with error bars.}
\label{fig3}
\end{figure}

We mention that the difference for larger $x_2$ values
between the $\beta=0$
case of Fig.\ 3 in the present work and Fig.\ 8 of 
ref.\ \cite{Dieperink97} is a 
consequence of the use of a more realistic dipole pion-nucleon-nucleon 
form-factor in the present calculation, and to a smaller extent
due to a different set of parton distributions \cite{Lai97}.

We remark that the small pion excess probability found in the
present work 
would also lead to a pion contribution to the EMC
ratio for $x\sim 0.1$ which is smaller than in some other
approaches \cite{Marco96}. 
However, this ratio is mostly sensitive
to the nucleon self-energy in the medium and the role of pions is
difficult to isolate.

We conclude: i) a simultaneous experiment on a neutron-rich
nucleus and an $N\approx Z$ nucleus with a 5\% accuracy should
in principle be able to isolate the see up-down asymmetry (term
proportional to $\beta$ in Eq.\ (\ref{dyratio2})); ii) about 
50\% of the $\bar u -\bar d$ difference in a nucleus comes from
nuclear effects. Therefore the assumption that nuclear effects 
are negligible as in
ref.\ \cite{Szczurek96} cannot be justified.

{\bf Acknowledgements}

We acknowledge useful discussions with S.\ Kumano.
This research was supported in part by an NWO (Netherlands)
fellowship and the
Hungarian Research Foundation (OTKA) grant T16594.
This work is also part of the research program of the 
"Stichting voor Fundamenteel Onderzoek der Materie" (FOM) with
financial support from the
"Nederlandse Organisatie voor Wetenschappelijk Onderzoek" (NWO).


\end{document}